\documentclass[aps,prd,showkeys,showpacs,superscriptaddress,floatfix]{revtex4-2}

\usepackage{graphics,epsfig}
\usepackage{epstopdf}
\usepackage{subfigure}
\usepackage{palatino}
\usepackage{changes}
\usepackage[colorlinks=true,linkcolor=blue,urlcolor=blue,citecolor=blue]{hyperref}
\usepackage[toc,page]{appendix}
\usepackage[normalem]{ulem}
\usepackage{adjustbox}
\usepackage{latexsym}
\usepackage{amsmath}
\usepackage{amssymb}
\usepackage{amsfonts}
\usepackage{times}
\usepackage{graphicx}
\usepackage{dcolumn}
\usepackage{amsmath}
\usepackage{array}
\usepackage{wrapfig}
\usepackage{multirow}
\usepackage{tabularx}
\begin{document}

\title{Weak deflection angle of Kazakov-Solodukhin black hole in plasma medium using Gauss-Bonnet theorem and its greybody bonding }

\author{Wajiha Javed}
\email{wajiha.javed@ue.edu.pk; wajihajaved84@yahoo.com} 
\affiliation{Division of Science and Technology, University of Education, 54770 Township-Lahore, Pakistan}

\author{Iqra Hussain}
\email{iqrahussain057@gmail.com} 
\affiliation{Division of Science and Technology, University of Education, 54770 Township-Lahore, Pakistan}

\author{Ali {\"O}vg{\"u}n}
\email{ali.ovgun@emu.edu.tr}
\affiliation{Physics Department, Eastern Mediterranean University, Famagusta, 99628 North
Cyprus via Mersin 10, Turkey.}

\date{\today}

\begin{abstract}
In this paper, we study light rays in a Kazakov-Solodukhin black hole. To this end, we use the optical geometry of the Kazakov-Solodukhin black hole within the Gauss-bonnet theorem. We first show the effect of the deformation parameter $a$ on the Gaussian optical curvature, and then we use the modern method popularized by Gibbons and Werner to calculate the weak deflection angle of light. Our calculations of deflection angle show how gravitational lensing is affected by the deformation parameter $a$. Moreover, we demonstrate the effect of a plasma medium on weak gravitational lensing by the Kazakov-Solodukhin black hole. We discuss that the increasing the deformation parameter $a$, will increase the weak deflection angle of the black hole. Our analysis also uncloak how one may find a observational evidence for a deformation parameter on the deflection angle. In addition, we studied the rigorous bounds of the Kazakov-Solodukhin black hole for the grey body factor and also studied the graphical behaviour of the bounds by fixing $l=0$ and $M=1$ and observed that increasing the deformation parameter $a$ will decrease the rigorous bound $T_b$ of the black hole.
  \end{abstract}



\maketitle

\section{Introduction}

 In 1915, Einstein cleverly came up with a new idea of theory of general relativity (GR) \cite{C1},
 that has been undergone through a number of experiments \cite{C3,C4,EventHorizonTelescope:2019dse}. It was founded that the experimental consequences and the conceptual predictions both were well fitted with each other. One of the greatest challenges in theoretical physics is the unification of quantum field theory and Einstein´s general relativity into a theory of quantum gravity. Using the known physical laws, the theories are not compatible with each other. However, to understand the inside of the Black hole and  the Big Bang, these theories should be combined and the quantum corrections must be located in the Schwarzschild solution according to the various approaches to unify them. Kazakov and Solodukhin show that the generalization of the Schwarzschild solution is possible by neglecting the non-spherical deformations using the effective scalar-tensor gravity \cite{Kazakov:1993ha} and its properties are studied in \cite{Konoplya:2019xmn,Peng:2020wun,Bezerra:2019qkx}. Furthermore gravitational lensing (GL) is one of the famous prediction and numerous individuals also verified these experiments \cite{C5,Dyson:1920cwa}. In 1801, GL was initially suggested by Soldner \cite{C6}. Gravitational lensing is a powerful tool to understand not only dark and massive objects \cite{C7,C8,C9} but it is also
 very effective to study black holes (BHs) and wormholes \cite{C10,C11,C12,Abdujabbarov:2017pfw,Chakrabarty:2018ase,Atamurotov:2015nra,Turimov:2018ttf,Virbhadra:1999nm,Virbhadra:2002ju,Virbhadra:2007kw,Virbhadra:2008ws,Keeton:1997by,Bozza:2002zj,Bozza:2009yw,Chen:2009eu,Sharif:2015qfa,Cao:2018lrd,Bisnovatyi-Kogan:2017kii,BisnovatyiKogan:2010ar,Cunha:2018acu,Cimdiker:2021cpz,Okyay:2021nnh,Ovgun:2021ttv}.

A BH is an astronomical object having such strong gravity that nothing could escape from it even electromagnetic radiations such as light \cite{C13}. The boundary of a BH from which nothing could escape is called event horizon (EH). Black hole behaves as an ideal black body because it reflects no light  \cite{C14,C15}. According to GR, the light beam is deflected by a small angle if it passes closely to mass $M$ with a huge effect parameter $b$ defined as \cite{C16,C24}:
\begin{equation}
   \gamma=\frac{2R_{S}}{ b}=\frac{4M}{b},G=c=1.
\end{equation}

In 2008, Gibbons and Werner (GW) founded another geometrical technique  to process the deflection angle  of light by the implementation of Gauss-Bonnet theorem (GBT) to the optical geometry  for asymptotically flat static BHs \cite{C25}.
Gibbons and Werner proclaimed that this formula gives the perfect result for Schwarzschild BH \cite{C25} in weak fields. Afterwards, this formula was enlarged to find the deflection angle of Kerr BH \cite{C27} and other objects such as rotating global monopole, cosmic string, wormholes and other solutions \cite{Ovgun:2019qzc,Crisnejo:2019ril,Ishihara:2016vdc,Jusufi:2017hed,Ovgun:2018ran,Jusufi:2017uhh,Jusufi:2017lsl,Ovgun:2019wej,Javed:2019ynm,Javed:2019kon,Sakalli:2017ewb,deLeon:2019qnp,Li:2019mqw,Jusufi:2017drg,Ovgun:2018fte,Ovgun:2018tua,448,Fu:2021akc,Kumaran:2019qqp,Kumaran:2021rgj,Kumar:2020hgm,Javed:2019rrg,Javed:2020lsg,Ishihara:2016sfv,  Ono:2017pie,Ono:2018jrv,OA2019,Arakida:2017hrm,Li:2020dln,Li:2020wvn,Javed:2020fli,Javed:2020fli2,Javed:2020mjb,ElMoumni:2020wrf,Pantig:2021zqe,Li:2019qyb,Li:2020zxi,Takizawa:2020egm,Gibbons:2015qja,Islam:2020xmy,Pantig:2020odu,Tsukamoto:2020uay,Jusufi:2017vta,Ovgun:2020yuv,Ovgun:2020gjz,Belhaj:2020rdb,Jusufi:2018jof,Javed:2020wsv,Ovgun:2018xys,Javed:2019qyg,Jusufi:2017mav,Jusufi:2018kmk,Ovgun:2018prw,OvgunUniverse,Ovgun:2018fnk,Li:2019vhp,Ono:2018ybw}. Currently, GW have changed the common point of view associated to the technique of calculating the deflection angle. By using GBT, one can obtain a beautiful expression to find the deflection angle.  Now, let shortly discuss the GBT also known as  Gibbons-Werner method that tie up the topologically surfaces and show that the deflection angle can be calculated as follows:
\begin{equation}
    \beta=-\int_{0}^{\pi} \int_{\frac{b}{r\sin\phi}}^{\infty} \mathcal{K} dS,\\
\end{equation}
where the $\mathcal{K}$ is the Gaussian optical curvature. The above equation is valid for the deflection angle of asymptotically flat spacetime, if we have non asymptotically flat metric then, one should use the
finite distance corrections \cite{Ishihara:2016vdc}. Ishihara et al. calculated the deflection angle of light for the finite distance (huge impact parameter) using the GBT when the source and the observer are at asymptotically flat region \cite{Ishihara:2016vdc}. Then, Crisnejo and Gallo \cite{Crisnejo:2019ril} use the GBT to obtain the gravitational deflections of light in a plasma medium in a black hole spacetime. Using this technique, more work was done to compute the deflection angle in plasma medium \cite{Javed:2020lsg}-\cite{C95}.\\
It was found that the null geodesics and the serious scattering issue associated with the rays impending from infinity together with the impact parameter are analogous with each other. S.Hawking manifest in his paper that BHs 
are grey in reality as the thermal radiations are emitted by the BHs  and are termed as Hawking Radiations (HR) \cite{C96}.
The BH emit radiations likewise the black body emit radiations at the EH although the initial radiations that were generated differ in the course of travelling through out the spacetime \cite{C97} thus the observer observing the radiations at infinity notice the distinct radiations as compared to the radiations at emitter that is so-called grey body factor (GBF) and is also considered to be sieve of HR \cite{C98,C99}.
The GBF can be computed by distinct techniques, but the most interesting technique is to determine the  rigrous bounds of the GBF \cite{C100,C101,C102}. 

The aim of this work is to calculate the deflection angle of the Kazakov-Solodukhin black hole \cite{Kazakov:1993ha,Konoplya:2019xmn,Peng:2020wun} by using the GBT. After that, we will contemplate the impact of plasma medium on the desired black hole and discuss  the effect of the deformation parameter $a$ on the deflection angle in vacuum and in plasma medium. In the last the bound of GBF will be computed and also be plotted  and studied in detail.

The work is formed as follows. In section 2, we first review the Kazakov-Solodukhin black hole. Then we calculate its optical geometry and the Gaussian optical curvature. In section 3, we devote this section  to find deflection angle of the Kazakov-Solodukhin black hole in weak field limits. In Section 4 we do the graphical analysis of the angle for non-plasma case. In section 5, we calculate the deflection angle in plasma medium. Furthermore, in section 6 we determine the graphical behavior of the deflection angle of the Kazakov-Solodukhin black hole. In section 7 we determine the bound of the GBF and in section 8 we study the lower bound graphically. We conclude in section 9.

\section{Kazakov-Solodukhin black hole }
 
 The Kazakov-Solodukhin black hole spacetime is described by the following line-element  \cite{Kazakov:1993ha}
\begin{equation}
 ds^2=-B(r)dt^2+ \frac{dr^2}{B(r)}+r^2 (d\theta^2+\sin^2\theta d\phi^2),\label{IQ1}
\end{equation}
with the metric function

\begin{equation}
    B(r)=\left(\frac{\sqrt{r^{2}-a^{2}}}{r}-\frac{2 M}{r}\right), \label{met1}
\end{equation}

and the asymptotic expression for the metric $r>>a$ large $r$ or small $a$ is
\begin{equation}
B(r)\approx 1-\frac{a^2}{2r^2}-\frac{2M}{r}, \label{metric}
\end{equation}

where BH mass is indicated by $M$ and $a$ is the deformation parameter. It is noted that the deformation parameter $a$ is typically to be identified with the Planck scale. Kazakov-Solodukhin black holes mimics the behavior of a charged classical Reissner-Nordström black holes. Note that there is a correspondence also between the Kazakov-Solodukhin black holes with the deformation parameter $a$ and the tidal charged black holes with the charge $q$. In the weak field limit, they would be deeply connected by the relation $q = -a^2/2$ and hence existence of a point charge leads to quantum fluctuations of the background manifold. \cite{Javed:2021arr}.
The surface gravity is calculated by
\begin{equation}
\kappa=
\left.\frac{1}{2} \frac{\partial B(r)}{\partial r}\right|_{r=r_{+}}
\end{equation}
and the Hawking temperature is found as follows \cite{Hajebrahimi:2020xvo}: \begin{equation}
T_{H}=\frac{1}{2 \pi} \frac{\sqrt{4 M^{2}+2 a^{2}}}{\left(2 M+\sqrt{4 M^{2}+2 a^{2}}\right)^{2}}.
\end{equation}
  Note that this black hole
radiate until it approaches the parameter $a$. Quantum deformation stops the 
black hole from completely evaporating \cite{Chen:2014jwq} similarly with the generalized uncertainty principle. Quantum deformation of the Kazakov-Solodukhin black hole results in a non-zero massive black hole 
remnant with finite temperature at the final stage of evaporation. This remnant might be a Planck-scale remnant, and can also be a potential candidate for dark matter \cite{Hajebrahimi:2020xvo}.

For static and spherically symmetric metric, the light source and
the observer are taken in the equatorial plane as $(\theta=\frac{\pi}{2})$. As we are working in null geodesics so, put $ds^{2}$=0 and we obtain the corresponding optical metric
\begin{equation}
 dt^2=\frac{dr^2}{(1-\frac{a^2}{2r^2}-\frac{2M}{r})^2}
 +\frac{r^2d\phi^2}{1-\frac{a^2}{2r^2}-\frac{2M}{r}}.\label{iq2}
\end{equation}
The Christoffel symbols other than zero using Eq. (\ref{iq2}) are as follow.
\begin{eqnarray}
\Gamma^{r}_{rr}&=&-\frac{{B}'(r)}{{B}(r)}.\nonumber\\
\Gamma^{r}_{\phi\phi}&=&-\frac{r^{2}{B}'(r)}{2}+r{ B}(r).\nonumber\\
and\\
\Gamma^{\phi}_{r\phi}&=&\frac{{B}'(r)}{2{B}(r)}-\frac{1}{r}.\nonumber\\
\end{eqnarray}
Then the Ricci scalar associated with the optical metric is computed as:
\begin{equation}
{R}={B}(r){B}''(r)-\frac{{B}'(r)^{2}}{2}.
\end{equation}
  The Gaussian optical curvature $\mathcal{K}$ can also be rewritten in:
\begin{equation}
\mathcal {K}=\frac{R_{r \phi r \phi}}{2},
\end{equation}
which yields for the optical metric of Kazakov-Solodukhin bh as:
\begin{equation}
\mathcal{K}\thickapprox-\frac{2M}{r^3}-\frac{3a^2}{2r^4}
-\frac{M^2}{r^5}+\frac{3Ma^2}{r^5}-\frac{a^4}{4r^6}.
\end{equation}

After simplifying, we can write it in weak field limits as follows:
\begin{equation}
 \mathcal{K}\thickapprox-\frac{2M}{r^3}-\frac{3a^2}{2r^4}+\frac{3Ma^2}{r^5}
+\mathcal{O}(M^2,a^4).\label{IQ3}
 \end{equation}

\section{Deflection angle of Kazakov-Solodukhin BH}
 In this section, we calculate the deflection angle of a Kazakov-Solodukhin bh using the GBT which provides relation between the intrinsic geometry of the spacetime and its topology of the region $\mathcal{A}_{R}$ with boundary $\partial A_{R}$, stated as \cite{C25}
\begin{equation}
 \int\int_{\mathcal{A}_{R}}\mathcal{K}dS+\oint_{\partial\mathcal{A}_{R}}kdt
 +\sum_{j}\epsilon_{j}=2\pi\mathcal{X}(\mathcal{A}_{R}),
\end{equation}

 where $\mathcal{K}$ is the Gaussian curvature and $k$ is the  geodesic curvature, stated as
 $k=\bar{h}(\nabla_{\dot{\gamma}}\dot{\gamma},\ddot{\gamma})$  such a way that $\bar{h}(\dot{\gamma},\dot{\gamma})=1$,
 here $\ddot{\gamma}$  represents
 unit acceleration vector and the $\epsilon_{j}$ is the exterior angle
 at the jth vertex. As $R\rightarrow\infty$, our two jump angles become $\pi/2$ or we can say that
 the sum  of the two jump angles to the observer and source satisfies $\theta_{O}+\theta_{S}\rightarrow\pi$.
  The Euler characteristic is $\mathcal{X}(\mathcal{A}_{R})=1$, as $\mathcal{A}_{R}$ is
 non singular. Consequently, we rewrite it as
\begin{equation}
 \int\int_{\mathcal{A}_{R}}\mathcal{K}dS+\oint_{\partial
 \mathcal{A}_{R}}kdt+\epsilon_{j}=2\pi\mathcal{X}(\mathcal{A}_{R}),
\end{equation}
 here, $\epsilon_{j}=\pi$ demonstrates that
 $\gamma_{\bar{h}}$ also the entire jump angle is a geodesic. Now, geodesic curvature $k(C_{R})=\mid\nabla_{\dot{C}_{R}}\dot{C}_{R}\mid$ should be calculated. Since, we can evaluate the
 geodesic curvature's radial part:
 
\begin{equation}
 (\nabla_{\dot{C}_{R}}\dot{C}_{R})^{r}=\dot{C}^{\phi}_{R}
 \partial_{\phi}\dot{C}^{r}_{R}+\Gamma^{r}_{\phi\phi}(\dot{C}^{\phi}_{R})^{2}.\label{IQ4}
\end{equation}

 For large $R$, $C_{R}:=r(\phi)=r=const$. The christofell symbols that are
 related to the optical geometry are given in the last equation   $\Gamma^{r}_{\phi\phi}$  . It is obvious from the above equation that initial term will vanish as topological effect is not involved and the second term will be calculated by using $k(C_{R})=\mid\nabla_{\dot{C}_{R}}\dot{C}_{R}\mid$. The geodesic curvature is obtained as follows:
\begin{equation}
 (\nabla_{\dot{C}^{r}_{R}}\dot{C}^{r}_{R})^{r}\rightarrow\frac{1}{R}.
\end{equation}
  Subsequently, we came up $ k(C_{R})dt=d\phi$. Then the GBT becomes
\begin{equation}
 \int\int_{\mathcal{A}_{R}}\mathcal{K}ds+\oint_{\partial \mathcal{A}_{R}} kdt
 =^{R \rightarrow\infty }\int\int_{S_{\infty}}\mathcal{K}dS+\int^{\pi+\beta}_{0}d\phi.\label{iqra2}
\end{equation}

 For calculating the deflection angle in the weak deflection limit at 0th order, we use the light beam which follows a straight line approximation and is defined as  $r(t)=b/\sin\phi$. The deflection angle is then obtained by using

\begin{equation}
 \beta=-\int^{\pi}_{0}\int^{\infty}_{b/\sin\phi}\mathcal{K}\sqrt{det{g}}drd\phi,\label{IQ5}
\end{equation}
 where
\begin{equation} \label{166}
 \mathcal{K}\sqrt{det{g}}=(\,{\frac {-15{M}^{2}{a}^{2}}{2{r}^{5}}}-3\,{\frac {{M}^{2}}{{r}^{3}}}
-3\,{\frac {{a}^{2}M}{{r}^{4}}}-2\,{\frac {M}{{r}^{2}}}-\,{\frac {3{
a}^{2}}{2{r}^{3}}}).
\end{equation}
 After using Eq. (\ref{166}) into Eq. (\ref{IQ5}), the deflection angle thus becomes
\begin{equation}
 \beta \thickapprox \frac{4M}{b}+\frac{3a^2\pi }{8b^{2}}+\frac{4a^2M}{3 b^{3}}+\mathcal{O}(M^2,a^4).\label{S1}
\end{equation}

Thus, one can immediately observe how the deformation parameter $a$ affect the weak deflection angle of the Kazakov-Solodukhin black hole. Such black holes may carry, distinctive from the Schwarzschild black holes, features that allow the observations of the remnant.

Weak and strong deflection
angle by quantum deformed Schwarzschild BH has already been investigated by X. Lu and Y. Xie \cite{Lu:2021htd} using different method. Moreover, they have calculated the deflection angle without plasma medium, only in vacuum. Before, obtaining the deflection angle they have expanded the metric function $B(r)$ and compared them with those of the re-normalization group improved Schwarzschild black hole \cite{Bonanno:2000ep} and the asymptotically safe black hole \cite{Held:2019xde}.\\
Then they have calculated the deflection angle of light $\hat\alpha$ for the given BH
by using \cite{Weinberg:1972kfs},
\begin{equation}
\hat\alpha=2\int^{\infty}_{r_0}\frac{dr}{r^2\sqrt{u^{-2}-B(r)r^{-2}}},
\end{equation}
where $B(r)=\left(\frac{\sqrt{r^{2}-a^{2}}}{r}-\frac{2 M}{r}\right)$ is the radial function of line-element and $u$ is representing
the impact parameter which is defined by
\begin{equation}
u=\frac{r_0}{\sqrt{B(r_0)}},
\end{equation}

In the above equations, $r_0$ is the closest approach between the photon
and the desired BH. The deflection angle they obtained is
\begin{align}
\hat\alpha(u)=\frac{4m_\bullet}{u} + \left(\frac{15}{4}+\frac{3 a^2}{8}\right)\pi\frac{m^{2}_\bullet}{u^2} + \frac{45 \pi}{64}\left(77+21a^2 + \frac{3 a^4}{4}\right)\frac{m^{4}_\bullet}{u^4}+\mathcal{O}\left(\frac{m^{5}_\bullet}{u^5}\right).\label{MA1}
\end{align}
While, in our analysis, we use Gibbons and Werner method using Gauss-Bonnet theorem for the calculation
of deflection angle $\beta$ for quantum corrected
Schwarzschild BH with the asymptotic expression of the metric given in \ref{metric} as $B(r)\approx 1-\frac{a^2}{2r^2}-\frac{2M}{r}$.  Moreover, it is to be noted that the angle (\ref{MA1}) obtained in the suggested paper contains more term of mass  $m_\bullet$ having higher orders, while in our analysis we only take the leading order  terms which contains BH mass $M$ of first order in the obtained angle (\ref{S1}). The results show the agreement between Gauss-Bonnet and geodesic approach breaks down for second order terms which is expected due to the straight line approximation used in Gauss-Bonnet method in the integration domain. Our calculations agree in leading order terms which was our main point. Moreover, we have also examined the deflection angle graphically with respect to the black hole mass, impact parameter and the deformation parameter.

\section{GRAPHICAL ANALYSIS for non-plasma medium} 
In this section, we study the graphical analysis of deflection angle on Kazakov-Solodukhin bh that has been computed in Eq. (\ref{S1}) . To check the effect of deformation parameter $a$, impact parameter $b$ and the mass $M$ on the deflection angle we likewise illustrate the deflection angle on plots. In the beginning we plot deflection angle $\beta$ in connection with the deformation parameter $a$ by variating the mass $M$ of the desired BH and the impact parameter $b$ and study its impact. Next, we plot  deflection angle $\beta$  with respect to $b$ and allocate distinct values to mass $M$ (large and small) and to the deformation parameter $a$ (small and large)  and study the consequences. At the end, deflection angle $\beta$ is plotted with the  deformation parameter $a$ by changing the values of the impact parameter and the deformation parameter and study the obtained results. The details of the above stated graphs and their results is discussed below.

\begin{center}
\epsfig{file=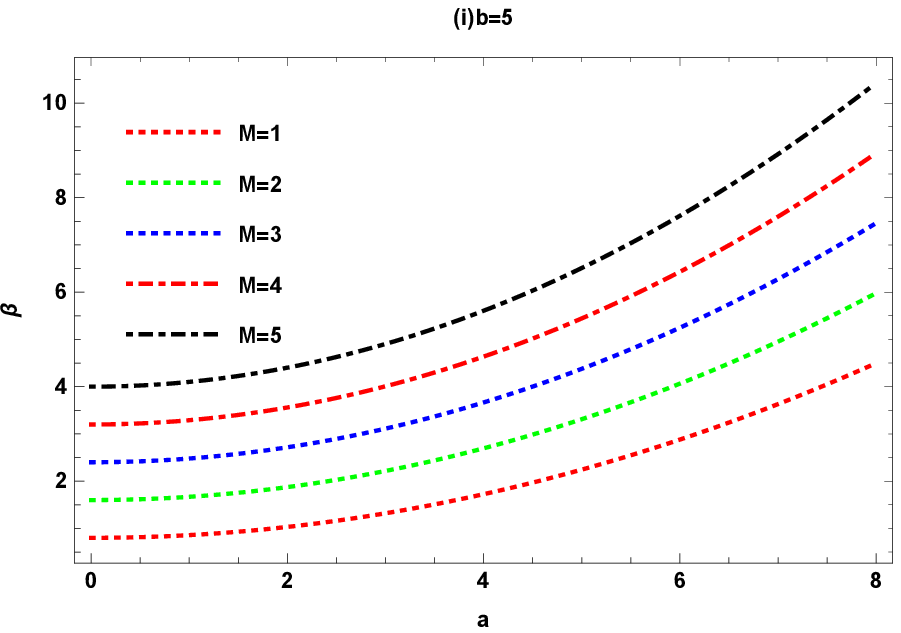,width=0.50\linewidth}\epsfig{file=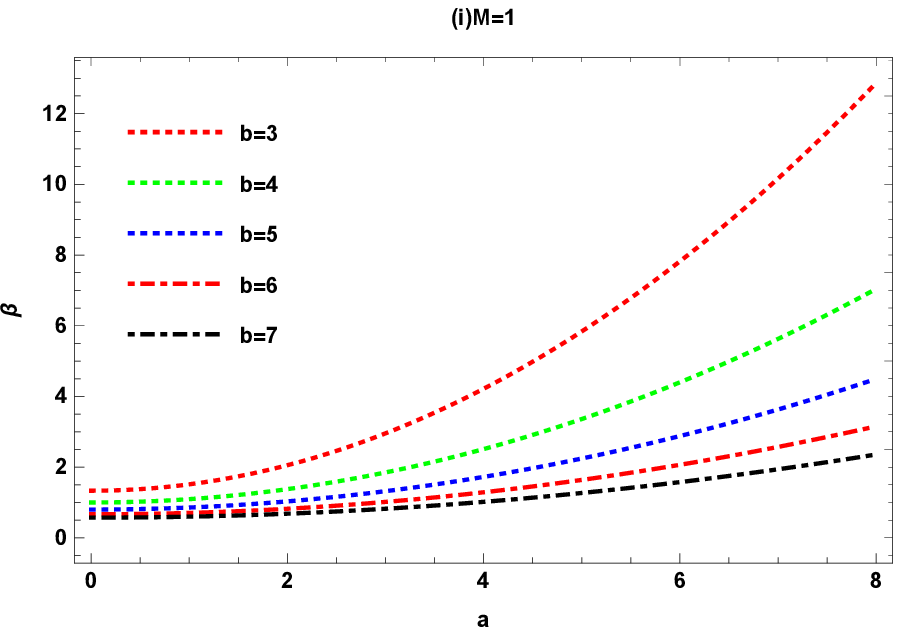,width=0.50\linewidth}\\
{Figure 1: Relation between $\beta$ and $a$}.
\end{center}
\begin{center}
\epsfig{file=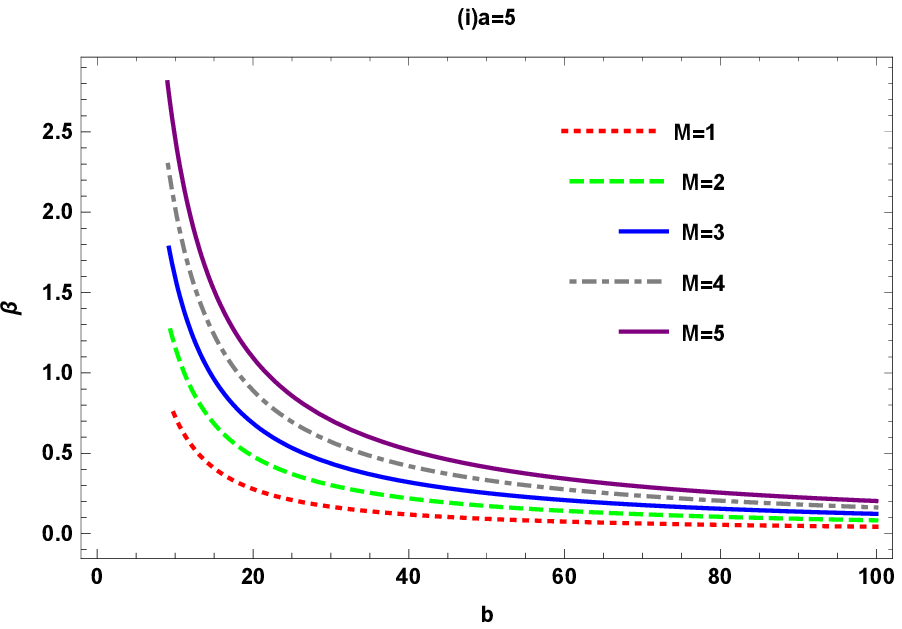,width=0.50\linewidth}\epsfig{file=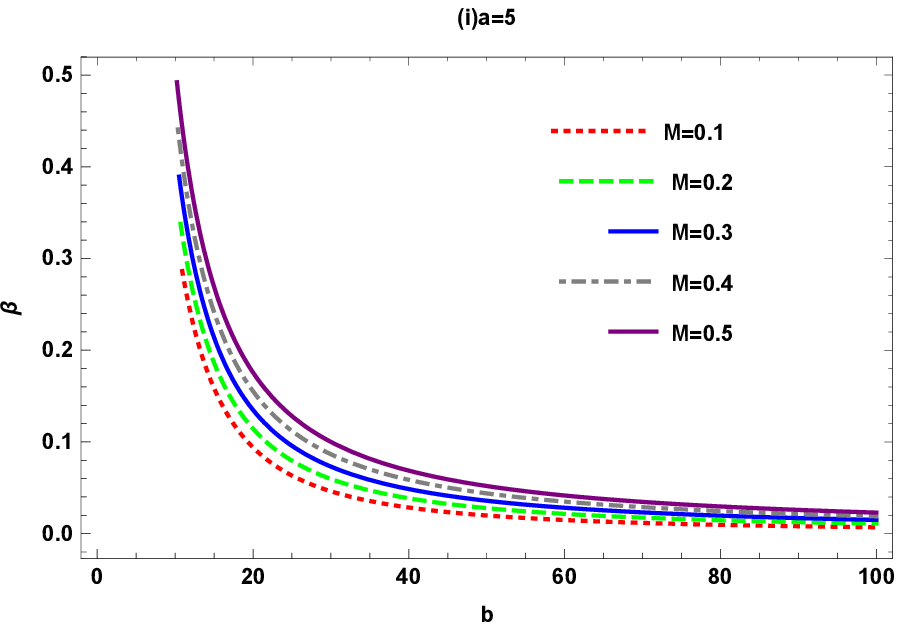,width=0.50\linewidth}\\
{Figure 2: Relation between $\beta$ and $b$}.
\end{center}
 \begin{center}
\epsfig{file=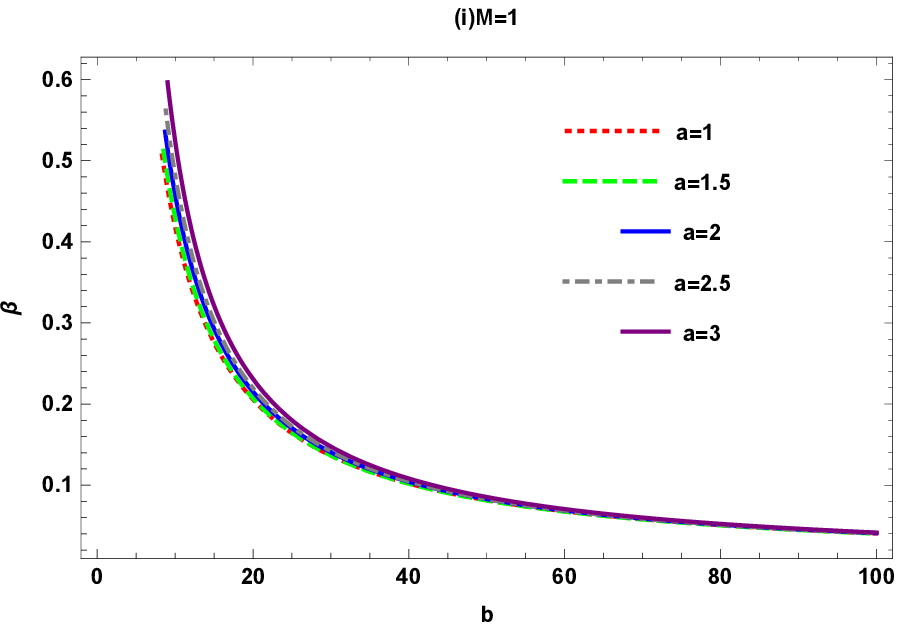,width=0.50\linewidth}\epsfig{file=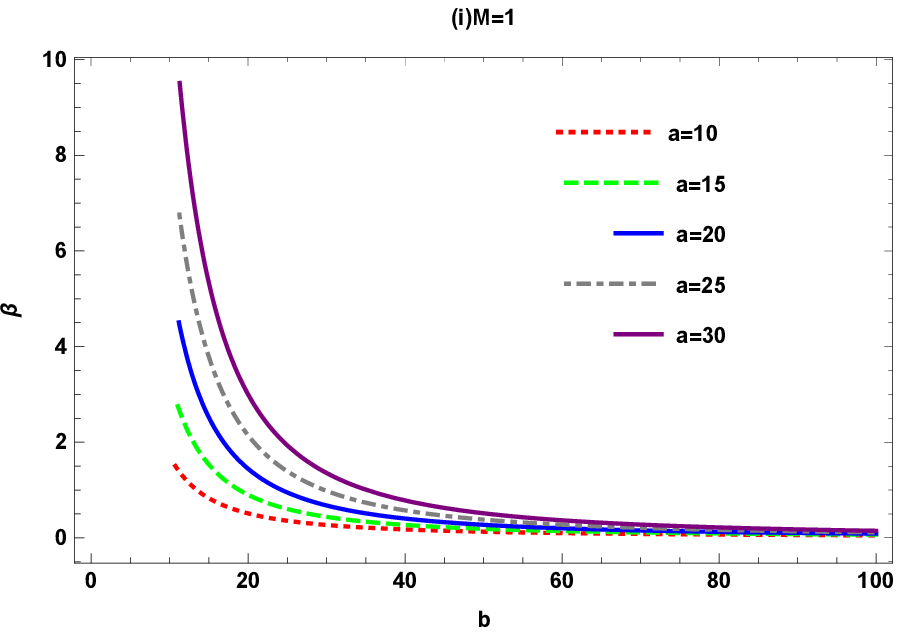,width=0.50\linewidth}\\
{Figure 3: Relation between $\beta$ and $b$}.
\end{center}
\begin{center}
\epsfig{file=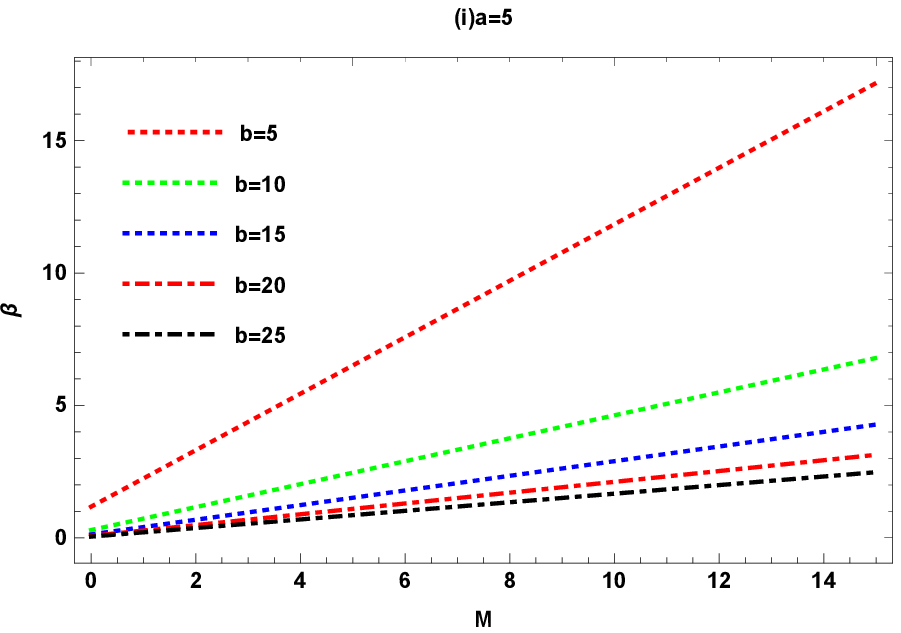,width=0.50\linewidth}\epsfig{file=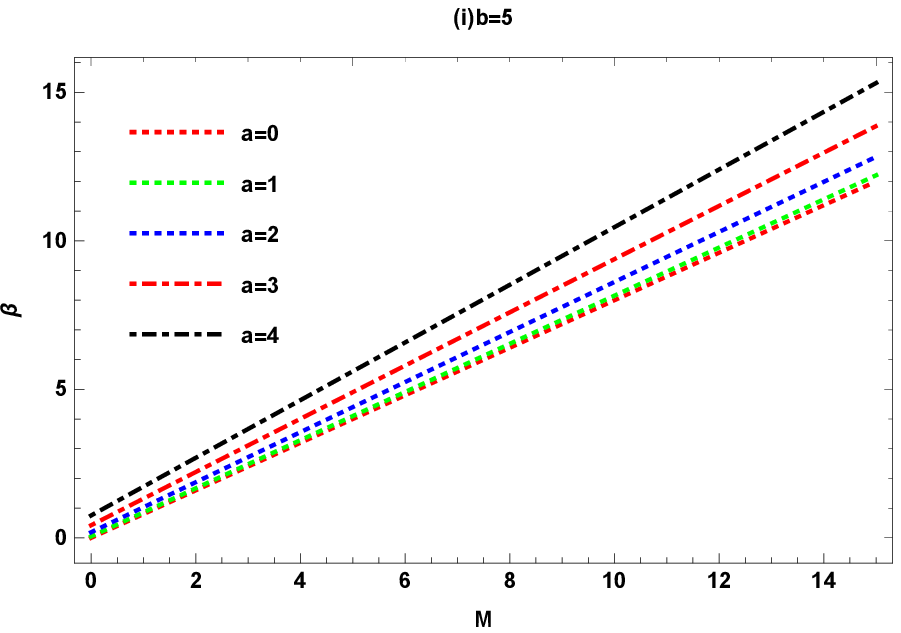,width=0.50\linewidth}\\
{Figure 4: Relation between $\beta$ and $M$}.
\end{center}
\subsection{Deflection angle $\beta$ w.r.t deformation parameter $a$}
\begin{itemize}
\item \textbf{Figure 1} This plot shows the behavior of $\beta$ w.r.t $a$ for different values of $M$ and $b$ and by selecting fix value of $b$ and $M$.
\begin{enumerate}
\item The left plot shows that the deflection angle is increasing uniformly for increasing the mass of the BH in which the impact parameter has a specific value $b=5$  and the domain of the deformation parameter is $0\leq{a}\leq 8$ .
\item  We show that in the right plot the deflection angle for the domain $0\leq{a}\leq 8$ and fixed value of the mass $M=1$ of the BH  is decreasing for the increasing  $b$.
\end{enumerate}
\end{itemize}
\subsection{Deflection angle $\beta$ w.r.t Impact parameter $b$}
\begin{itemize}
\item \textbf{Figure 2} shows the behavior of $\beta$ w.r.t $b$ by fixing the value of the deformation parameter $a=5$  and varying $M$ and the domain of $b$ is considered to be $0\leq {b} \leq 100$ .
\begin{enumerate}
\item In the left plot, it can be  noticed that the deflection angle $\beta$ is gradually increasing for the large values of $M$.
\item In the right plot, it can be observed that the deflection angle $\beta$ increase for  small  variation of $M$.\\
  Therefore, it can be concluded that the deflection angle increases in both the cases (for large and small
    values of $M$).
    \end{enumerate}
\end{itemize}
\begin{itemize}
\item \textbf{Figure 3} shows the behaviour of deflection angle w.r.t impact parameter by keeping the value of $M$ fixed that is $M=1$ and giving variation to the value of deformation parameter $a$ .
\begin{enumerate}
\item In both the left and the right plots, it can be noticed that $\beta$ gradually increases in both the cases (for small and large variation of $a$).
 \end{enumerate}
\end{itemize}
\subsection{Deflection angle $\beta$ w.r.t  $M$}
\begin{itemize}
\item \textbf{Figure 4} shows the behavior of $\beta$ w.r.t $M$ by fixing $a=5$ and for large variation of $b$ and for fixed value of $b=5$ and for small values of $a$. The domain of mass is assumed to be $0\leq {M} \leq 15$ respectively.
    \begin{enumerate}
\item In the left plot, we noticed  $\beta$ decreases for the  small variations of $b$ which exhibits the inverse relation of the attained deflection angle and the impact parameter.
\item In the right  plot, we noticed  $\beta$ is increases for small variations of $a$ which shows the direct realtion of the angle and the deformation parameter.
 \end{enumerate}
\end{itemize}
\section{Gravitational lensing by Kazakov-Solodukhin bh in plasma medium} 
 In this section, we study the effect of plasma medium on the gravitational lensing of a Kazakov-Solodukhin bh. For this purpose, we consider the corresponding BH in plasma having refractive index $n$,
\begin{equation}
 n^2\left(r,\omega(r)\right)=1-\frac{\omega_e^2(r)}{\omega_\infty^2(r)}.
\end{equation}
 The refractive index reads
\begin{equation}
 n(r)=\sqrt{{1-\frac{\omega_e^2}{\omega_\infty^2}\left(1 -\frac{a^2}{2r^2}-\frac{2M}{r}\right)}}.
\end{equation}
 
  Similarly as before, we obtain the corresponding optical metric as follows:
\begin{equation}
 dt^2=g^{opt}_{xy}dx^xdx^y=n^2 \left[\frac{dr^2}{B^2(r)}+\frac{r^2d\phi^2}{B(r)}\right],\label{iqra3}
\end{equation}
the determinant  ($g^{opt}_{xy}$) of the above optical metric is defined as follows:
\begin{equation}
 \sqrt{g^{opt_{xy}}}=r(1-\frac{\omega_e^2}{\omega_\infty^2})+M(3
 -\frac{\omega_e^2}{\omega_\infty^2})+\frac{a^2}{4r}(3
 -\frac{\omega_e^2}{\omega_\infty^2}).
\end{equation}
 By using Eq. (\ref{iqra3}), we can find the following three non-zero Christofell symbols as
\begin{equation}
    \Gamma^r_{rr}=(1+\frac{\omega_e^2 B}{\omega_\infty^2})\left[-B^\prime B^{-1}(1-\frac{\omega_e^2 B}{\omega_\infty^2})-\frac{B^\prime \omega_e^2}{2\omega_\infty^2}\right],\nonumber
\end{equation}
\begin{equation}
    \Gamma_{\phi r}^{\phi}=(1+\frac{\omega_e^2B}{\omega_\infty^2})\left[r^{-1}(1-\frac{\omega_e^2B}{\omega_\infty^2}-\frac{B^\prime B^{-1}}{2}(1-\frac{\omega_e^2B}{\omega_\infty^2})-\frac{B^\prime \omega_e^2}{2\omega_\infty^2}\right]\nonumber
\end{equation}
 and
\begin{equation}
    \Gamma^r_{\phi \phi}=(1+\frac{B\omega_e^2}{\omega_\infty^2})\left[-rB(1-\frac{B\omega_e^2}{\omega_\infty^2})
    +\frac{r^2B^\prime}{2}(1-\frac{B\omega_e^2}{\omega_\infty^2})+\frac{r^2B}{2}\frac{B^\prime\omega_e^2}{\omega_\infty^2}\right].\nonumber
\end{equation}
 Gaussian optical curvature in terms of curvature tensor can be stated as
\begin{equation}
    \mathcal{K}=\frac{R_{r\phi r\phi}(g^{opt_{xy}})}{det(g^{opt_{xy}})},\label{iqra1}
\end{equation}
 by the use of Eq. (\ref{iqra1}) in the weak field limit Gaussian optical curvature is written as:
\begin{eqnarray} \label{KK}
\mathcal{K}&\thickapprox&12\,{\frac {{M}^{2}{\omega_e}^{2}}{{\omega_\infty}^{2}{r}^{
4}}}+3\,{\frac {{M}^{2}}{{r}^{4}}}-3\,{\frac {M{\omega_e}^{2}
}{{\omega_\infty}^{2}{r}^{3}}}-2\,{\frac {M}{{r}^{3}}}-16\,{\frac {
{M}^{2}{a}^{2}{\omega_e}^{2}}{{\omega_\infty}^{2}{r}^{6}}}+
13\,{\frac {M{\omega_e}^{2}{a}^{2}}{{\omega_\infty}^{2}{r}^
{5}}}+3\,{\frac {M{a}^{2}}{{r}^{5}}}-\,{\frac {5{a}^{2}{\omega_e}^{2}}{2{\omega_\infty}^{2}{r}^{4}}}-\,{\frac {3{a}^{2}}{2{r}^{
4}}}.
\end{eqnarray}

For calculating angle in the weak field limits of the light rays, we follow a straight line approximation so that we can utilize the condition  $ r=\frac{b}{sin\phi}$ within the GBT
\begin{equation} \label{pl}
    \beta=-\lim_{R\rightarrow 0}\int_{0} ^{\pi} \int_\frac{b}{\sin\phi} ^{\infty} \mathcal{K} dS.
\end{equation}

 By using Eq. (\ref{KK}) within the Eq. (\ref{pl}), the deflection angle $\beta$ in plasma medium is found as follows:
\begin{eqnarray}
  \beta&\thickapprox&\,4\,{\frac {M}{b}}+ \left( {\frac {3\,\pi}{8\,{b}^{2}}}+{\frac {4\,M}{3
\,{b}^{3}}} \right) {a}^{2}+ \left( 2\,{\frac {M}{b{\omega_\infty}^
{2}}}+ \left( {\frac {\pi}{4\,{b}^{2}{\omega_\infty}^{2}}}-{\frac {
M}{{b}^{3}{\omega_\infty}^{2}}} \right) {a}^{2} \right) {\omega_{e}}^{2}
\label{S3}
\end{eqnarray}

It is shown that the photon rays move in a
medium of homogeneous plasma and taking
$(\frac{\omega_e}{\omega_\infty}\rightarrow0)$, Eq. (\ref{S3}) reduces to Eq. (\ref{S1}), and the effect of the plasma is removed.

\section{GRAPHICAL ANALYSIS for Plasma medium}
In this section, we study the graphical behavior of deflection angle $\beta$ in the plasma medium. Moreover, we also overview the physical significance of these plots to investigate the impact of plasma medium as well as the deformation parameter $a$. Here, we consider $\frac{\omega_e}{\omega_\infty}$=$10^{-1}$ and give variation to the impact parameter $b$, deformation parameter $a$ and to the mass $M$ in order to acquire these plots. These plot were same as the plots obtained for the non-plasma medium.
\begin{center}
\epsfig{file=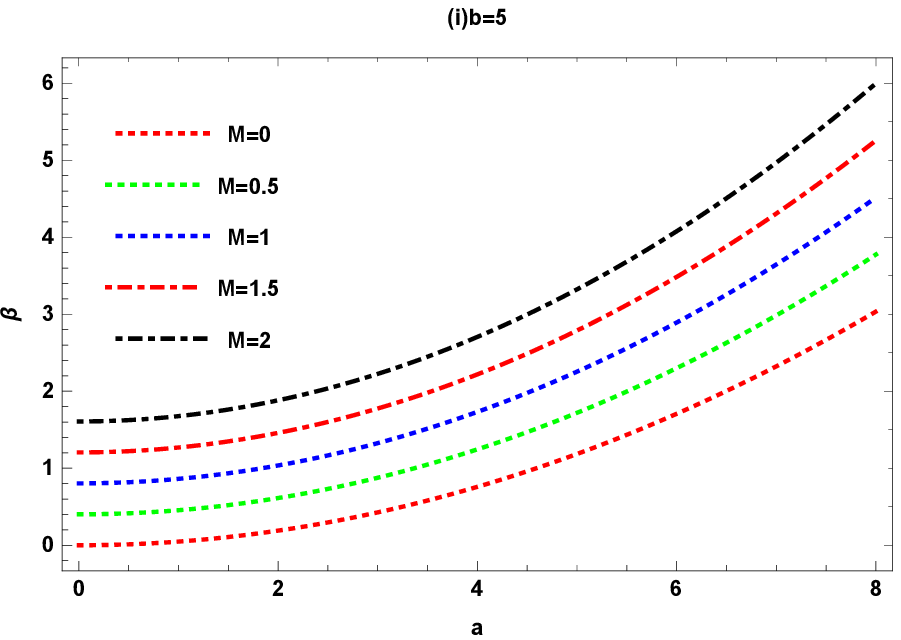,width=0.50\linewidth}\epsfig{file=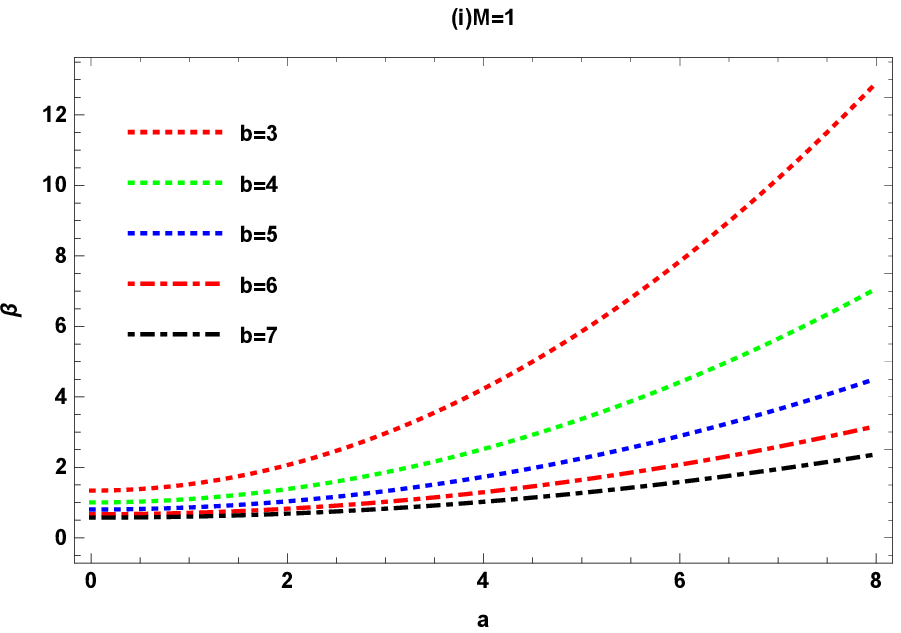,width=0.50\linewidth}\\
{Figure 5: Relation between $\beta$ and $a$}.
\end{center}
\begin{center}
\epsfig{file=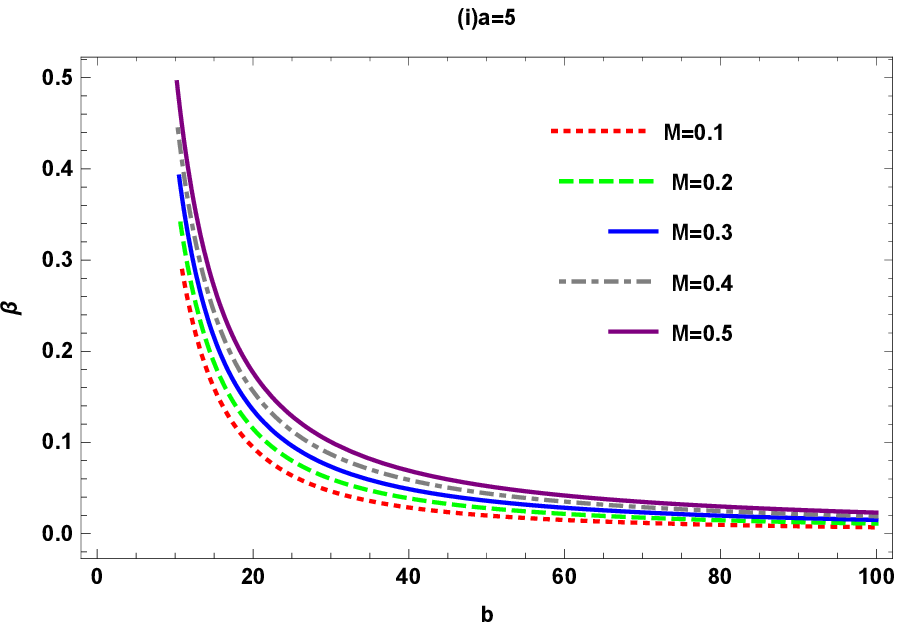,width=0.50\linewidth}\epsfig{file=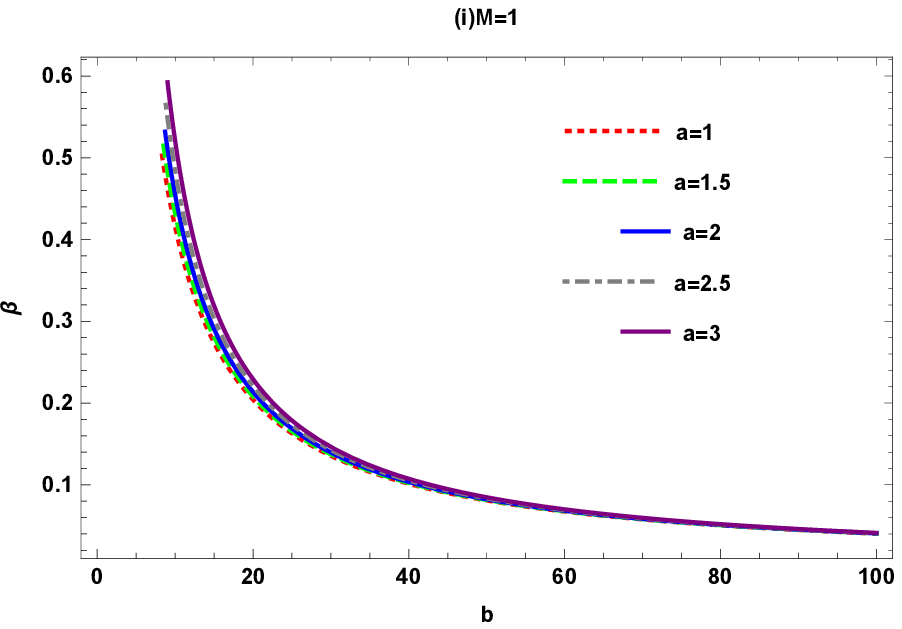,width=0.50\linewidth}\\
{Figure 6: Relation between $\beta$ and $b$}.
\end{center}
\begin{center}
\epsfig{file=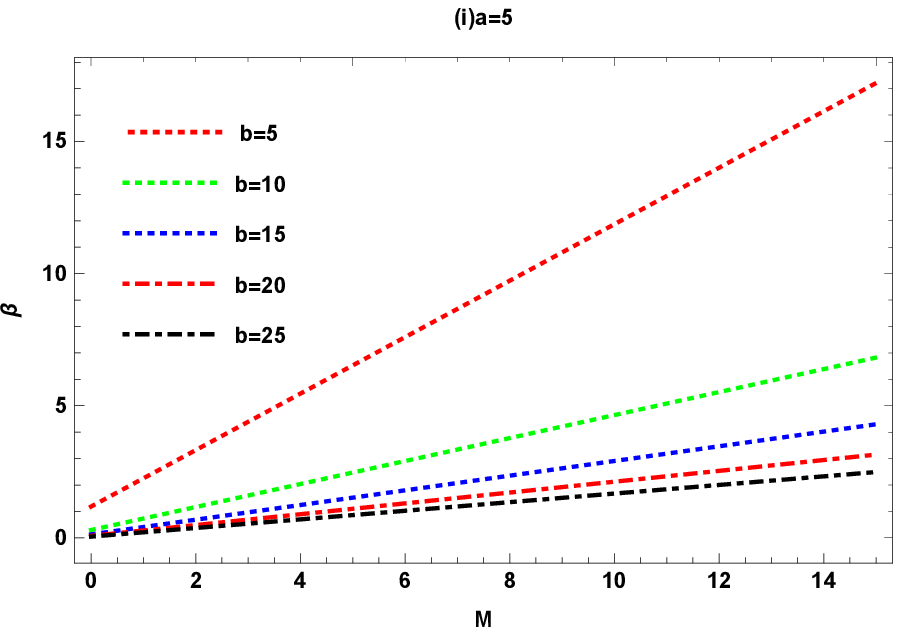,width=0.50\linewidth}\epsfig{file=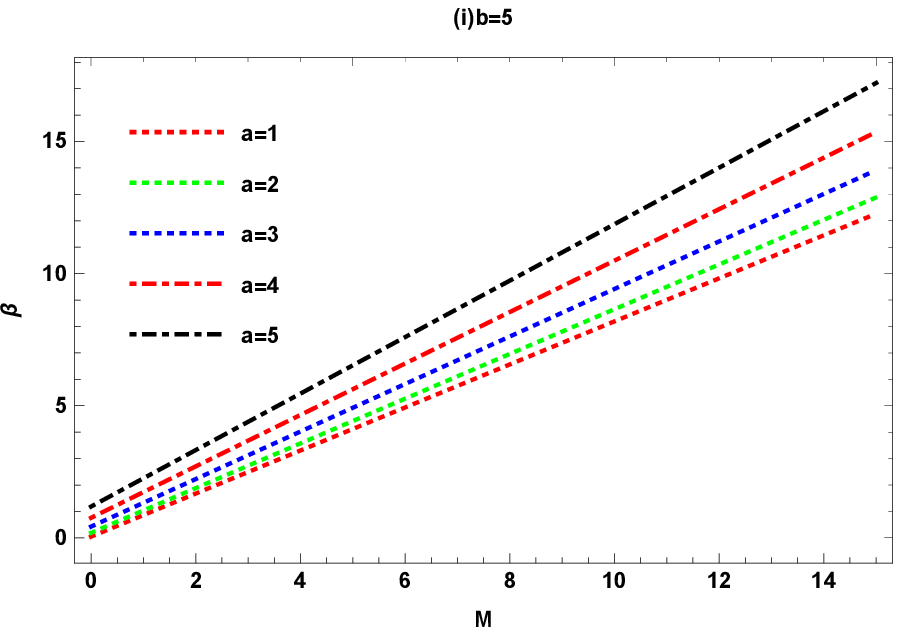,width=0.50\linewidth}\\
{Figure 7: Relation between $\beta$ and $M$}.
\end{center}
\subsection{Deflection angle $\beta$ w.r.t  $a$}
\begin{itemize}
\item \textbf{Figure 5} displays the behavior of $\beta$ w.r.t $a$ for fix values of $b=5$ and for varying the values of $M$ and for fix values of $M=1$ and for varying the values of $b$ respectively.
\begin{enumerate}
\item In the left plot,  we observe that the deflection angle gradually increases for small variation of $M$ and the domain of the deformation parameter is taken to be $0\leq {a} \leq 8$ .
\item In the right plot, we observed that deflection angle exponentially decreases for large variation of impact parameter and for specified $M=1$ which shows the inverse relation of the impact parameter and the deflection angle.
    \end{enumerate}
\end{itemize}
\subsection{Deflection angle $\beta$ w.r.t  $b$}
\begin{itemize}
\item \textbf{Figure 6} shows the behavior of $\beta$ w.r.t $b$ for fix values of $a=5$ and for varying the values of $M$ and for fix values of $M=1$ and for varying the values of $a$ respectively. Here the domain of impact parameter is $0\leq {b} \leq 100$
\begin{enumerate}
\item In the left plot, we observed that deflection angle $\beta$  is gradually increases for small variation of $M$ and approaches to positive infinity.
\item In the right plot, we observed that deflection angle $\beta$ is gradually increases for huge variation of $a$ and approaches to positive infinity.\\
In both the situations the angle is exhibiting the direct relation.
 \end{enumerate}
\end{itemize}
\subsection{Deflection angle $\beta$ w.r.t  $M$}
\begin{itemize}
\item \textbf{Figure 7} demonstrate the behavior of deflection angle $\beta$ w.r.t $M$ by fixing the values of deformation parameter $a=5$ and by varying the values of impact parameter $b$ and  fixing the values of deformation parameter $b=5$ and by varying the values of impact parameter $a$ respectively
\begin{enumerate}
\item In the left plot, we observed that deflection angle $\beta$ exponentially decreases for large variations of impact parameter  $b$  and deflection angle $\beta$ rapidly decrease for $5<b<10$.
\item In the right plot, we observed that deflection angle gradually increases for small variations of deformation parameter $a$.
 \end{enumerate}
\end{itemize}

Weak deflection angle of tidal charged black hole (TCBH) in plasma medium has already been
worked out by Javed, Hamza and Ovgun \cite{Javed:2021arr}. They have obtained the angle in terms of
the mass, impact parameter and the tidal charge of the BH. Mainly, they have studied
the BH under the influence of tidal charge. In graphical analysis, they have plotted the
angle of deflection with respect to the impact parameter for different values of the tidal charge. 
It was observed that by increasing the charge the obtained angle is decreasing and the impact 
of tidal charge is different in non-plasma and plasma mediums.

Black holes endowed with more intense (negative) tidal charges present bigger absorption 
cross sections, while their scattering spectra in the
weak field regime is not easily distinguishable from that of Schwarzschild black hole. 
Tidal charges allow black holes to spin faster than extreme Kerr black hole. 
Tidal charge increases the efficiency of energy extraction from the black hole 
via the Penrose process \cite{Khan:2019gco}. The size of the event horizon plays also an 
important role in the quantity of energy which can be extracted from the black hole. 
Tidal charge also changes the event horizon size, the increase of the tidal charge 
intensity may result in an increase or decrease of maximum amplification obtained in the scattering process.

Recently, in our analysis, we obtain the deflection angle in terms of deformation parameter,
impact parameter and the mass of the BH. In the graphical conduct,
we take variation of the impact parameter, mass and the deformation parameter
of the BH and study the outcomes obtained from these variations in detail.
We observe that when we plot the graph between the angle of deflection and
the impact parameter by changing the values of the deformation parameter then
the angle is gradually increasing and approaches to positive infinity.
Similarly, when we plot the graph between the obtained angle and the
mass of the BH by changing the values of the deformation parameter
then the angle is again increasing and exhibiting the positive behaviour. 
Moreover, the effects of the deformation parameter on graphical analysis 
are same in non-plasma and plasma mediums.

In gravitational lensing the angle of deflection of photon is independent of frequency
in non-plasma medium \cite{Bisnovatyi-Kogan:2015dxa}. While, for the case of plasma medium, the frequency of photon is depending upon the
radial coordinate $r$ because of the existence of the gravitational field \cite{Bisnovatyi-Kogan:2015dxa}. We can observe that
in gravitational lensing numerous images of the same source are formed \cite{Bisnovatyi-Kogan:2017kii} as
exhibited in the section of the graphical conduct of the paper. It is fascinating that as
exhibited in non-plasma case the graphical conduct of deformation parameter is similar to
the graphical conduct of the deformation parameter in plasma medium having frequency. It
is also studied that the photon having larger deformation parameter bends larger. The
impact of the deformation paramater on the angle exhibits physical behaviour regardless
of the medium. Thus, the observational effects of the obtained angle and the
corresponding graphical behaviour for various parameters prove that the obtained results are physical will hold for astrophysical tests. 

In 1960's, the refraction on gravitation was investigated
using the propagation of radio signals in the solar corona where the light rays oscillating near to the Sun, are supposed to be deflected due to the solar gravity and by plasma in the solar corona and its corresponding deflection angles were calculated by Muhleman
et al \cite{Muhleman:1966,Muhleman:1970zz,teu}. This kind of deflections are known as chromatic (depending on the frequency of the photon). These kind effect of plasma medium might be observed in the low frequency radio observations as stated in \cite{Bisnovatyi-Kogan:2017kii}.

\section{Determination of greybody bound of Kazakov-Solodukhin BH}
This section is based on the calculations needed for the determination of the rigrous  bound of the GBF of the Kazakov-Solodukhin BH. 
We can utilize different techniques like WKB technique or other matching methods. The most analytical technique is to derive the bound of GBF and the bound can be stated as \cite{C96,C97}\\
\begin{equation}
    T\geq sech^2\left(\frac{l}{2\omega}\int_{-\infty} ^{\infty}{\mathcal{V}(r)}dr_*\right)\label{T1}.
\end{equation}
The Kazakov-Solodukhin metric in D dimension can be stated as, \cite{Kazakov:1993ha}\\
\begin{equation}
 ds^2=-B(r)dt^2+ \frac{dr^2}{B(r)}+r^2 (d\theta^2+\sin^2\theta d\phi^2),\nonumber\\
\end{equation}
with the metric function

\begin{equation}
    B(r)=\left(\frac{\sqrt{r^{2}-a^{2}}}{r}-\frac{2 M}{r}\right),
\end{equation}
The two event horizons(interior and exterior) that are calculated can be written as,
\begin{equation}
 r_+={\sqrt{4M^{2}+a^{2}}}
\end{equation}
\begin{equation}
    r_-=-{\sqrt{4M^{2}+a^{2}}}.
\end{equation}
The Schrodinger like equation attains the following form
\begin{equation}
    \left[\dfrac{d^2}{dr_*^2}+\omega^2-\mathcal{V}(r)\right]\psi=0
\end{equation}
and here $dr_*=\frac{1}{B(r)}dr$\\
$\mathcal{V}(r)$ in the above equation is called the potential and is expressed as
\begin{equation}
   \mathcal{V}(r)=\frac{(d-2)(d-4)}{4}\frac{B^2(r)}{r^2}+\frac{(d-2)}{2}\frac{B(r)\partial_rB(r)}{r}+l(l+d-3)\frac{B(r)}{r^2}
\end{equation}
Now we will compute lower bound in which $h$ is a positive fcn and satisfies $h(-\infty)=h(\infty)=\omega$.
so, the bound given in Eq.(\ref{T1}) takes the form
\begin{equation}
    T_b\geq sech^2\left(\frac{1}{2\omega}\int_{r_+} ^{\infty}\frac{\mathcal{V}(r)}{B(r)}dr_*\right)
\end{equation}
Then
\begin{equation}
    T_b=sech^2\left(\frac{1}{2\omega}\int_{r_+} ^{\infty}(\frac{(d-2)(d-4)}{4}\frac{B(r)}{r^2}+\frac{(d-2)}{2}\frac{\partial_rB(r)}{r}+\frac{l(l+d-3)}{r^2})dr\right)
\end{equation}
Afterthat when we take d=4 then the above equation becomes
\begin{equation}
    T_b=sech^2\left(\frac{1}{2\omega}\int_{r_+} ^{\infty}(\frac{\partial_rB(r)}{r}+\frac{l(l+1)}{r^2})dr\right)
\end{equation}
later when the integral is solved and the value of the obtained ${r_+}$ is substituted we obtain,
\begin{equation}
    T_b\geq sech[\frac{\frac{l(l+1)}{\sqrt{4M^{2}+a^{2}}}+\frac{\pi}{4a}-\frac{ArcTan[\frac{2M}{a}]}{2a}}{2\omega}]^2
\end{equation}
Hence we have computed the expression for the lower bound of the Kazakov-Solodukhin bh. When we put $a^2=0$ the computed lower bound reduces to schwarzschild lower bound. \\
\section{GRAPHICAL CONDUCT}
After obtainning the equation for the lower bound of GBF of the desired BH we plot the bound with respect to $\omega$  and in each graph we allocate a fixed value to $l=0$ and in every graph we fix $M=1$ and assign different values to the deformation parameter. We assign both the positive values and the negative values to the deformation parameter of the Kazakov-Solodukhin bh. Here, in the plots, we also changed  the 
domain of $\omega$.
\begin{center}
\epsfig{file=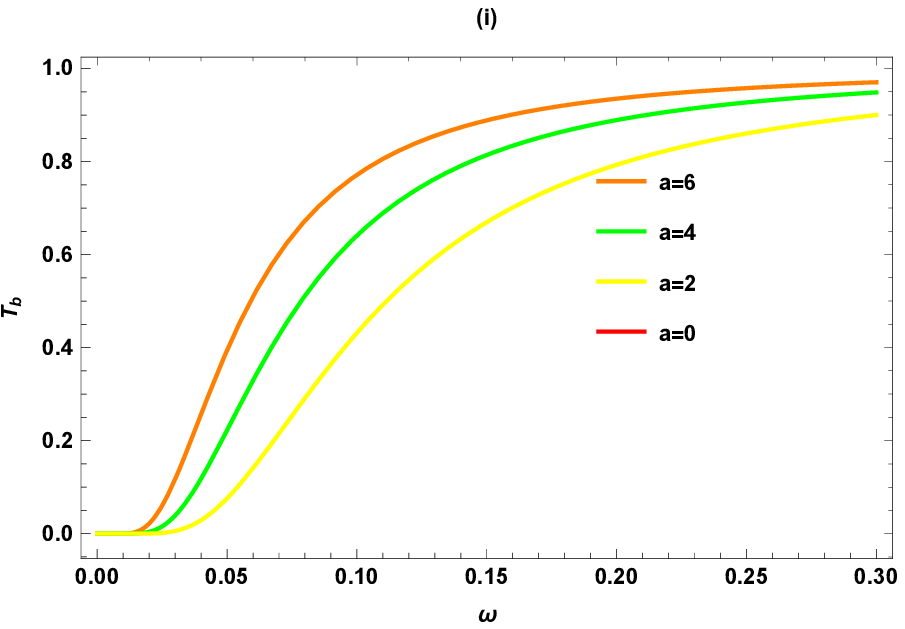,width=0.50\linewidth}\\
{Figure A: Relation between $T_b$ and $\omega$}.
\end{center}
\begin{center}
\epsfig{file=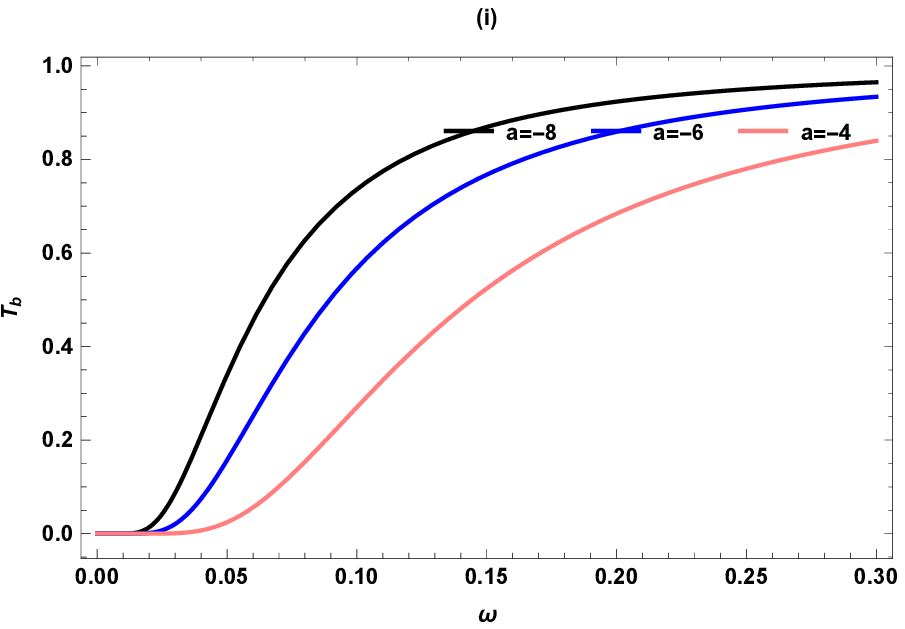,width=0.50\linewidth}\\
{Figure B: Relation between $T_b$ and $\omega$}.
\end{center}
\begin{center}
\epsfig{file=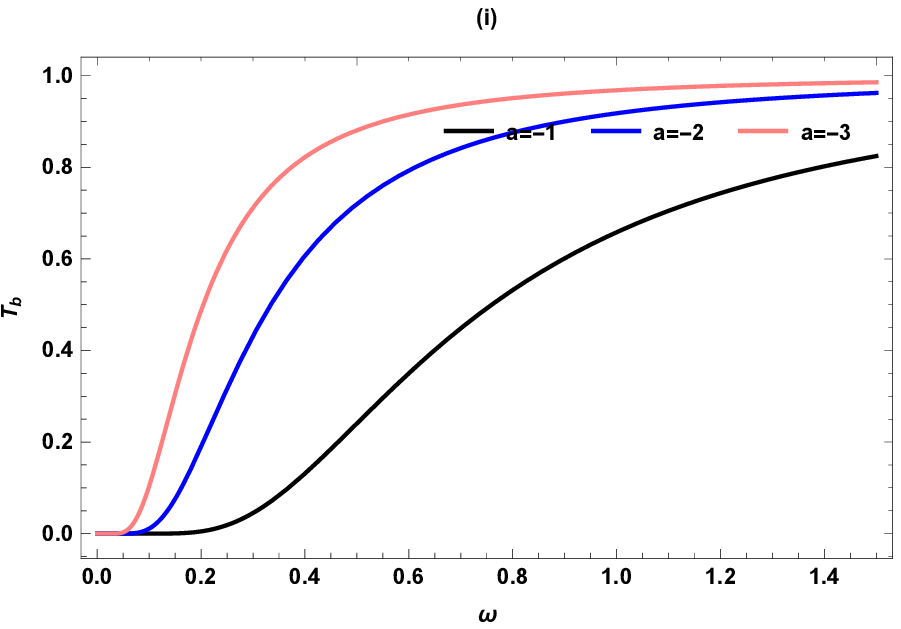,width=0.50\linewidth}\\
{Figure C: Relation between $T_b$ and $\omega$}.
\end{center}
\subsection{lower bound $T_b$ w.r.t $\omega$ for $l=0$ }
\begin{enumerate}
\item In figure A, we perceived that for assigning large variation to the deformation parameter and selecting the domain of $\omega$ 
to be $0\leq \omega\leq 0.30$  the computed lower bound of the Kazakov-Solodukhin bh is decreasing . This behaviour 
indicates the inverse relation of the obtained $T_b$ and the $a$ of the bh.
 \end{enumerate}
 \begin{enumerate}
\item In figure B, we recognized that the calculated bound $T_b$ when plotted against $\omega$ and negative values are given to the deformation parameter of the bh is declining.\\
Hence in both circumstances (+ive and -ive values of $a$) the calculated bound $T_b$ is exhibiting the same behaviour.
 \end{enumerate}
 \begin{enumerate}
\item In figure C, the domain of $\omega$ is selected to be $0\leq \omega\leq 1.5$ and negative terms are assign to the deformation parameter. It is seen that the bound is again approaching to the lower 
side of the plot.\\
In all the plots the obtained bound $T_b$ is showing the similar behaviour.\\
It is concluded  that higher the value of the deformation parameter $a$ the larger the potential will be and this results in the lowering of the $T_b$ of GBF.
\end{enumerate}
\section{Summary}
 In this paper, we have obtained the deflection angle for Kazakov-Solodukhin bh in both plasma and non-plasma medium. First, we have used a method proposed by GW known as GBT (a new geometric technique), to calculate the deflection angle for the desired BH. We have obtained the deflection angle $\beta$ of light by integrating a domain area outside the impact parameter  $b$, which shows that gravitational lensing is a global effect and is a good tool to investigate the features of the BH. In our analysis, we have shown the effect of the deformation parameter $a$ on the deflection angle $\beta$ in the weak field limit. Hence, the computed deflection angle (\ref{S1}) is as follow
 
\begin{equation}
      \beta \thickapprox \frac{4M}{b}+\frac{3a^2\pi }{8b^{2}}+\frac{4a^2M}{3 b^{3}}+\mathcal{O}(M^2,a^4).
\end{equation}

 It is noted that if we select $a=0$ in the above equation, the calculated deflection angle reduces to the Schwarzschild deflection angle up to the first order term. Furthermore, we likewise study the graphical analysis of deflection angle for the Kazakov-Solodukhin bh. Moreover, by considering the homogeneous plasma medium, we have shown that the deflection angle of light is (\ref{S3}):
 
\begin{eqnarray}
     \beta_{plasma}&\thickapprox&\,4\,{\frac {M}{b}}+ \left( {\frac {3\,\pi}{8\,{b}^{2}}}+{\frac {4\,M}{3
\,{b}^{3}}} \right) {a}^{2}+ \left( 2\,{\frac {M}{b{\omega_\infty}^
{2}}}+ \left( {\frac {\pi}{4\,{b}^{2}{\omega_\infty}^{2}}}-{\frac {
M}{{b}^{3}{\omega_\infty}^{2}}} \right) {a}^{2} \right) {\omega_{e}}^{2}.
\end{eqnarray}

Now, in order to remove the plasma effect, we neglect this term $(\frac{\omega_e}{\omega_\infty}\rightarrow0)$ and observe that the deflection angle in plasma medium reduces into vacuum results.

We have also studied the graphical analysis of deflection angle w.r.t impact
parameter b, deformation parameter $a$.
The outcomes achieved from the influence of deflection angle
stated in the paper are summed up as follows:
 Our main aim is to discuss the behavior of deflection angle for different values of impact parameter, deformation parameter and to the mass M.\\
\textit{\textbf{Deflection angle w.r.t Impact parameter:}}
 \begin{enumerate}
 \item   We have noticed that deflection angle is gradually increasing for small variation of $M$.
\item It is also noted that the deflection angle is gradually increasing for tiny variation of $a$.
\end{enumerate}
 \textit{\textbf{Deflection angle w.r.t Deformation parameter:}}
 \begin{enumerate}
\item Here, it is is noticed that deflection angle gradually increasing for small variation of $M$.
\item It is also noticed that that deflection angle is exponentially decreasing for the variation of $b$.
\end{enumerate}
\textit{\textbf{Deflection angle w.r.t M:}}
 \begin{enumerate}
\item It is observed that deflection angle is exponentially decreasing for large variation of impact parameter as follows:

\begin{equation}
 \beta \thickapprox \frac{4M}{b}+\frac{3a^2\pi }{8b^{2}}+\frac{4a^2M}{3 b^{3}}+\mathcal{O}(M^2,a^4).
\end{equation}

\item It is also examined that deflection angle is gradually increasing for small variation of deformation parameter $a$.
The quantum deformation starts to affect the deflection angle from the second-order term . Such black holes may carry, distinctive from the Schwarzschild black holes, features that allow the observations of the remnant \cite{Peng:2020wun,Lu:2021htd}. Moreover, Lu and Xie give some bound on the quantum deformation parameter as $0<a<1.53$ based on the apparent shadow size of the $M87^*$ observed in  EHT \cite{Lu:2021htd}.

After that, we worked out the the lower bound of the GBF and as a result obtained the following expression for the grey body bound of the Kazakov-Solodukhin bh.\\ 
\begin{equation}
    T_b\geq sech[\frac{\frac{l(l+1)}{\sqrt{4M^{2}+a^{2}}}+\frac{\pi}{4a}-\frac{ArcTan[\frac{2M}{a}]}{2a}}{2\omega}]^2
\end{equation}
At the end, we study the graphical behaviour of the bound in connection with $\omega$ by varying the deformation parameter. It was found that the attained bound decreased when we variate deformation parameter that shows the inverse relation of the $ T_b$ and the deformation parameter $a$. 

In the future, the astrophysical observations might shed light on the effect of deformation parameter on the deflection angle. A small remnants by the observations of the deflection angles can also give us hints on the dark matter. Hence, any discovery of the
deformation parameter would be an important signal beyond the general relativity.

\end{enumerate}

\begin{acknowledgments}	
The authors are grateful to the anonymous referees for their valuable comments and suggestions to improve the paper. The datasets generated during and/or analysed during the current study are available from the corresponding author on reasonable request.
\end{acknowledgments}

\end{document}